# Information Security Risks Assessment: A Case Study

SAMUEL CRIS AYO, BONAVENTURE NGALA, OLASUNKANMI AMZAT,

ROBIN LAL KHOSHI, SAMARAPPULIGE ISURU MADUSANKA

{u1815973, u1637893, u1709040, u1740368, u1634440}@uel.ac.uk

ABSTRACT

Owing to recorded incidents of Information technology inclined organisations failing to respond effectively to threat incidents, this project outlines the benefits of conducting a comprehensive risk assessment which would aid proficiency in responding to potential threats. The ultimate goal is primarily to identify, quantify and control the key threats that are detrimental to achieving business objectives.

This project carries out a detailed risk assessment for a case study organisation. It includes a comprehensive literature review analysing several professional views on pressing issues in Information security. In the risk register, five prominent assets were identified in respect to their owners. The work is followed by a qualitative analysis methodology to determine the magnitude of the potential threats and vulnerabilities. Collating these parameters enabled the valuation of individual risk per asset, per threat and vulnerability.

Evaluating a risk appetite aided in prioritising and determining acceptable risks. From the analysis, it was deduced that human being posed the greatest Information security risk through intentional/ unintentional human error. In conclusion, effective control techniques based on defence in-depth were devised to mitigate the impact of the identified risks from risk register.

## 1. INTRODUCTION

To make sure information or data stored in computers are not at risks of being compromised or tampered with, organisations need to deploy Information security management techniques. A risk is an effect of uncertainty on objectives and risk management as the coordinated activities to direct and control an organisation with regard to risk [1]. Risk management process is the systematic application of management policies, procedures and practices to the activities of communicating, consulting, establishing the context and identifying, analysing, evaluating, treating, monitoring and reviewing risk. Risk assessment is the overall process of risk identification, risk analysis and risk evaluation [1] [2]. Risk management maximises organisation output while minimising chances of unexpected outcomes [3].

This brings us to security management and why organisations should ensure information security management systems are in place. Information security management systems preserve the confidentiality, integrity and availability [4] of information by applying a risk management process and gives confidence to interested parties that risks are adequately managed. For this report, SparTax Collection Agency (SCA), a medium-sized sub-Sahara African enterprise is the organisation this report will focus on because African economies are the most vulnerable to cyber-attacks [5].

Africa is facing several Internet-related challenges concerning security risk, intellectual property infringement and protection of personal data [6] which has, for instance, cost South African economy $573 million, the Nigerian economy $200 million, and the Kenyan economy $36 million [6]. Small and Medium-sized Enterprises (SMEs) account for over 95% of firms and 60%-70% of employment and generate a large share of new jobs in Organisation for Economic Co-operation and Development (OECD) economies [7]. Similar trends are being observed globally, and risks brought about by disruptive internet technologies such as the information security breaches. Because SCA is also an SME just like the majority of other firms globally, the same techniques applied in this report can be applied across several organisations globally.

SparTax Collection Agency (SCA) is an organisation contracted by the Finance Authority of one of the Sub-Sahara African local governments to collect revenue in the form of taxes. The organisation has to be viewed as that which uses taxpayers' information with integrity to meet local government revenue obligations for it to be viewed as that which focuses on the developmental interests of the local government. SCA deploys a Revenue Management System (RMS) for all taxpayers account information and revenue management. The RMS is a one-stop platform where all data is integrated and analysed. It is also integrated online to other third-party systems and tools of partner agencies and institutions such as ports clearance systems, standards clearance system among others for coordinated day-to-day operations. In addition, SCA has contracted three other companies (Lamogi+, A&A and PAK) to provide other services of identifying new taxable businesses, maintaining and monitoring transit cargo real-time. The agency has also contracted a third-party company *MyCloud* to provide additional cloud services as a backup to their data centre.

The agency employs 1550 staff, 100 of which belong to the Information Technology (IT) and Information Security (IS) department. Among them are 10% of the staff is responsible for Information Security Management (ISM) in the agency while the rest with various levels of undocumented security clearances are charged with other IT support services within the company. More than half of SCA's staffs are on outsourced contract basis or internship and are on a minimum wage payment plan, which has continuously been an issue of complaint among many staffs who feel it is unfair. *Fig.1* below shows the company's management organisational structure. All the activities coordination at SCA are dependent on the support staff at their respective workstations exchanging communications by email, phone calls or SMS alerts over the public network infrastructure for incidents management and control.

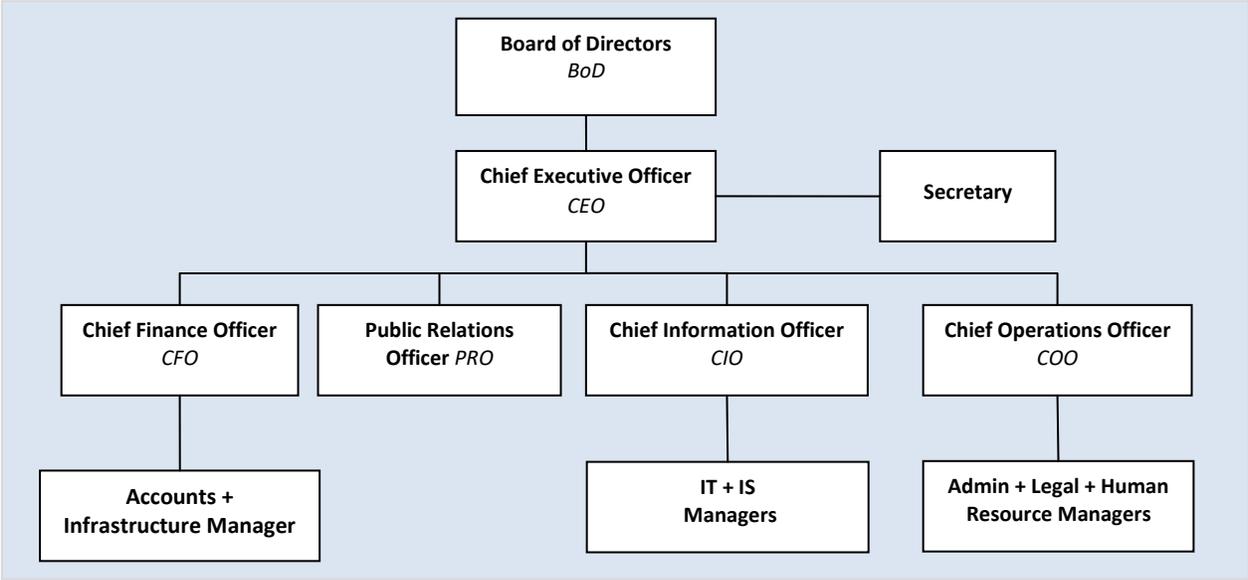

Fig. 1. SparTax Collection Agency (SCA) Organisational Structure

The company has had one general-purpose information security management policy for the last five years, but only department managers and selected staff members have been adequately trained on the policy. Implementation of the policy across the company has never been of paramount significance. However, since June 2016, the agency has had three major security breaches among other minor attempts leading to some undisclosed financial losses to the company. It is on this background that SCA has been tasked to develop a clear and thorough ISM strategy and report to be presented to the stakeholders.

In Section 2 below, we shall review five publications relating to cybersecurity, data breach impacts, economic valuation for Information Security investment, assessing Information Security value in organisations and Information Security Investment Decisions making. Then under Section 3, we shall describe the scope of our assessment, identify critical assets for the organisation, their threats and vulnerabilities and proceed to analyse the results tabulated in the risk register regarding risk appetite, choice and impact of controls. The conclusion of final work is presented in the last section.

## 2. LITERATURE REVIEW

**2.1** Schatz et al analysed several existing definitions given to cybersecurity [8]. The approach was used to derive an appropriate definition which was deemed fit based on several factors. These involved identifying the predominant and acceptable definitions, expressing viewpoints from the contentious differences and finally concluding on the best match through specific techniques. From the literature review conducted, viewpoints from several distinct references where weighed against each other and it could be observed that the difference lied between the scope of the concept, motive, opportunity and form of attack. A series of tools and techniques were utilised to compare and extract notable similarities amongst top definitions from selected sources. Techniques used involved inclusion and exclusion criteria which involved refining the search for definitions based on a series of best-fit conditions. A text mining framework tool was also used to examine keyword similarities before which dataset were normalised to standardise the character encoding. A stemming process was also carried out to reduce the number of distinct word types and increase the frequency of occurrence of significant word types. This rigorous process resulted in attributing values to the selected definitions to create an avenue for comparison. The aim was to fish out the most relevant attributes from the definition pool. Finally, a co-occurrence network analysis method was used to craft a new definition using a Minimum Spanning Tree (MST) network graph model in addition to filtering the term frequency to disintegrate the information to the salient points further. The derived definition was proven to be representative regarding semantic similarity and not just from a human-derived perspective. It was validated through the same semantic analysis benchmarking by weighing it up against the existing definition, and it ranked tops in this analysis.

Just as with any improvement, infallibility cannot be guaranteed, a significant limitation and area of improvement of this paper has been identified. In my opinion, utilising techniques to determine key similarities from a set of definitions which has arguably been subjectively deemed viable is not the ultimate approach to derive the most qualitative and relevant definition. There is a possibility that the selected definitions are not optimum in existence regarding relevance to what cybersecurity genuinely mean. Although the probability could be very slim, this research could potentially be a situation of picking out a decent definition from a bad bunch. It is believed that further analysis of more explicit definitions (including definitions defined in foreign languages) could yield a better-improved definition. Also, as pointed out by Schatz, Bashroush and Wall [8], the new derived definition stemmed from an iteration of sentences constructed manually, better still automatic means could be capable of iterating more combinations leading to a more relevant definition. Nevertheless, the proposed definition has proved to be unbiased and explicit given that it takes vital points from various viable sources.

**2.2** According to Schatz and Bashroush [9], the impact of repeated data breach events on organisations' market value, examines the influence of information security breaches on organisations stock market value. The study adopted the event study methodology, whereby data breaches for businesses category data was obtained from Privacy Rights Clearinghouse (PRC), analysed and the findings projected to the broader economic impact of such events. However, the quality and quantity of data obtained from PRC was not sufficient and mainly focused on the United States of America (USA) based entities as acknowledged by Schatz and Bashroush [9]. The result, therefore, gave a weak conclusion that organisations stock prize suffers from a publicly announced information security breach.

Irrespective of the sample size, data breaches affect performance and reputation of organisations. Therefore, organisations and investors should take more initiatives of having secure and continually managed and maintained information systems. Also, [9] focuses on business breach category and does not factor in government breach category which although most of them are not publicly traded in the stock market, their reputation, a key asset and to that matter "value" is affected by breaches.

**2.3** Economic Valuation for Information Security Investment: A Systematic Literate Review [10] addresses the issue of economic evaluation for information security investment. Daniel Schatz and Rabih Bashroush have conducted a systematic literature review on methods used by organisations to evaluate considered critical elements of decisions regarding information security investments. They pinpoint out that even if there are researches conducted to evaluate suitable approaches measuring expenditure of information security, there is a knowledge gap in practitioners to identify key criteria regarding information security investment. The contributions are a comparison framework, a catalogue of existing approaches; trends that help navigate existing work. Daniel Schatz and Rabih Bashroush point out that there is no tangible return on investment and further explains how information security measures aim to reduce loss. Therefore, it involves identification of current and future threats on assets, identification of highest valued assets and implementation of controls to reduce or mitigate risks. It further points out challenges faces by security professionals transferring risks into financial formulas. The Authors present several arguments presented by various research papers on multiple approaches taken by organisations.

The authors have conducted a systematic literature review (SLR) to provide structured and systematic methods to search, examine and evaluate current research questions. It follows the guidance provided by Kitchnham and Charters [11], Brereton et al. [12], Biochini et al. [13] and Cronin et al. [14]. The authors started the SLR process with the construction and definition of five research questions. The authors have constructed a search mechanism to capture relevant material which accommodates defined research questions. The authors have conducted systematic searches in several databases for research papers. The search scope narrowed on relevance to computer science and information security field. The results were filtered based on inclusion and exclusion criteria and database refinements, and it reduced down the results to 270 from 779.

Further reductions have resulted in 25 papers and used for data extractions. Authors have extracted nine important approach categories. After analysing the results, three main categories were chosen. They are Return on Investment (ROI), Real Options Theory (ROT), and Utility Maximization (UM).

**2.4** This study [15] describes a tested model that key constructs to consider when assessing the value of organisation information security assets. The research focuses on practitioners and researchers in IT security field and try to contribute significant knowledge on information security value chains in an organisation. The authors have proposed an evidence-based model. It combines theoretical work with real-world scenarios for assessing information security values in an organisation. The proposed model comprises five lament variables namely drivers, threats, accounting aspects, business environment and security capabilities [16]. They represent the critical areas of the context, how these lament variables relate to each other was investigated and were analysed. Then it was examined which relationship is significant using the Partial Least Square – Structural Equation Modelling (PLS-SEM) [17]. PLS-SEM model enhances support for the proposed model by authors. To verify the proposed model, authors have collected a high volume of data and analysed using a survey instrument specially designed using the structured equation modelling (SEM). The authors have emphasised that ignorance of relevant threats could have a negative impact on the organisation and it has a positive correlation with the security maturity of the environment. Security capabilities are therefore significant in accomplishing overall business values.

**2.5** The study by Daniel and Rabih [18] follows a grounded theory approach focused on information security investments and handling by experienced practitioners which are condensed into 15 principles. It highlights current practices, the key drivers and challenges in security investment strategies as well as organisational business environment factors relevant. Notable in the study is that decisions for Information security investments processes are initiated by "driving factors" and adjusted by "Challenges and Constraints". From this study, it's found that when information security investments support businesses to safely innovate, increase market agility, and enhance customer trust, they become competitive as compared to conventional budgeting approaches where funds are directed towards a 'minimum protection/maximum compliance' strategy rather than initiatives that contribute the most value to the organisation. The study illustrates the need to have a decision support process along with evidence-based approaches to avoid having problems. It encourages the adoption of standards such as ISO/IEC 27001 to identify control areas, not in line with control frameworks and ensures investments add value to an organisation.  When it came to security control effectiveness metrics, it was scarce and so the need to address the shortfall. Controls that could add overall value to an organisation at comparable costs had multiple benefits. In the study, practitioners are encouraged to adopt economic value approaches for financial valuations or performance models to justify security investments. Failure to adopt these values should relegate them to compliance and audit function. Eventually, risk reduction metrics are to be used to measure Information security program values.

The study [18] brought out the keys investment decision principles but faces a challenge in the diversity scope which makes the finding partial. The study is however limited to the United Kingdom and a small portion in the USA, all of which are developed economies compared to the rest of the developing economies. Since information security is becoming a global issue, the study should be expanded to include the rest of the developing countries as well, and the criteria for selecting practitioners should be elaborated more clearly.

## 3. RESULTS AND DISCUSSION

### 3.1 Purpose

Undertaking the risk assessment is to strategically identify the prominent threats and vulnerabilities that pose to hinder the organisation from achieving its business objective. The result from the risk register would be utilised in devising appropriate mitigation plans to reduce identified risk level to the barest minimum.

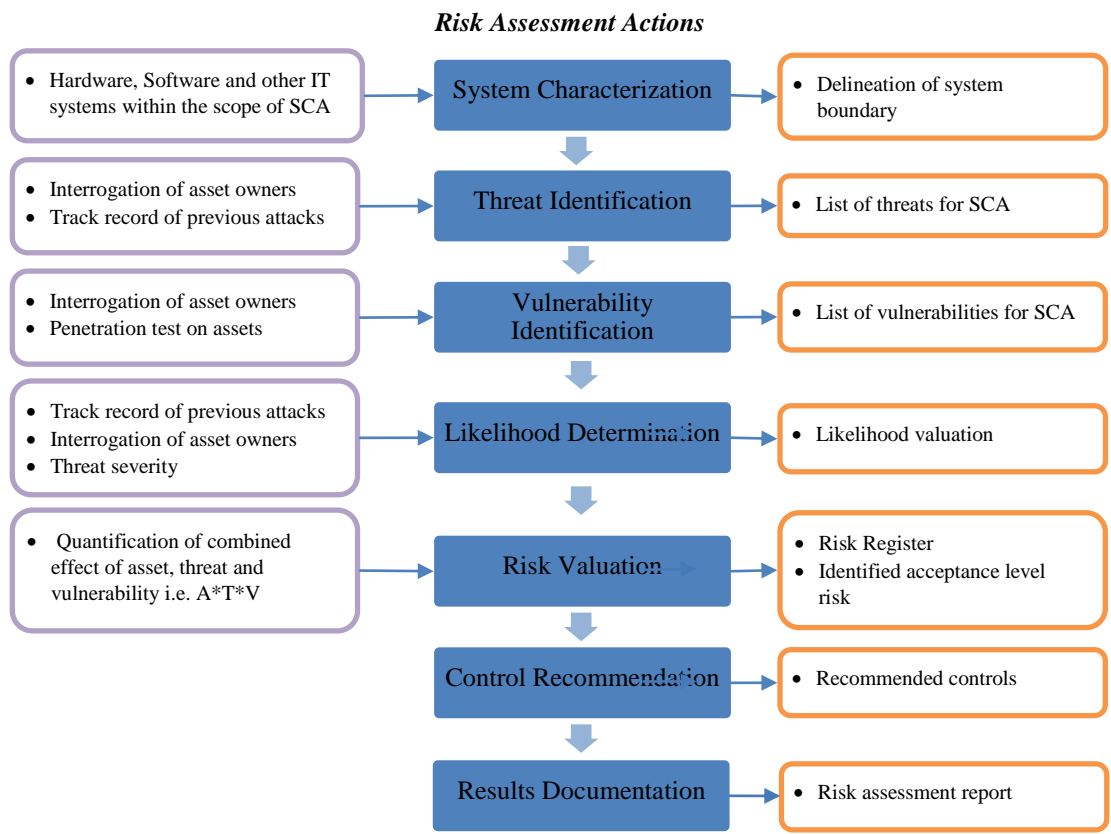

Fig . 2. Risk Assessment Flowchart (Adopted and modified from [19])

## 3.2 Risk assessment methodology flowchart

The *Fig.2* above gives a diagrammatic representation of the risk assessment methodology flowchart. It includes an orderly step-by-step process of the stages involved in carrying out the risk assessment for the organisation.

## 3.3 Scope and assets identified

After project sizing, we examine ISM of the significant assets to SCA according to management requirements. As stated under the Introduction Section, SCA is a model organisation depicting the general trends in the ISM in sub-Sahara Africa. We have followed a structural process to identify the organisation assets and their values using *TABLE.1*, then identify and collect similar threats and vulnerabilities from the respective asset owners.
Eventually, we quantify the likelihood and impact of these threats with a commonly adopted industry model as elaborated later below.

| Asset Values | Type of effect level of effect | Company embarrassment level | Personal safety implication | Personal privacy infringement | Failure to meet legal obligations | Financial loss (£) | Disruption to activities (£) (time & effort to recover from incident) |
|---|---|---|---|---|---|---|---|
| 1 | **Insignificant** | Contained within Work Area at worst | Minor injury to individual | Isolated personal detail revealed | Civil suit resulting in less than £10k damages | Up to 10k | Up to 10k |
| 2 | **Minor** | Contained within Company at worst | Minor injury to several people | Isolated personal detail compromised | Civil suit (above £10k). Small fine (up to £1k) | 10k to 100k | 10k to 100k |
| 3 | **Significant** | Local public or Press become aware | Major injury to individual | Several personal details revealed | Large fine (above £10k) | 100k to 500k | 100k to 500k |
| 4 | **Major** | National public or Press become aware | Major injury to several people or death of individual | Several personal details compromised | Custodial sentence imposed | 500k to 1000k | 500k to 1000k |
| 5 | **Acute** | Senior Staff forced to resign or Company fails | Death of several people | All personal details revealed and/or compromised | Multiple civil or criminal suits | Above 1000k | Above 1000k |

TABLE.1 Asset Values using Impacts of Incidents Matrix

On review, justifications and approval from management of the completed project sizing, we summarized the top five valuable assets as Pure Information in the form of Electronic data (data at rest, data in use and data in motion) without which SCA's business operations would be close to impossible, Physical IT hardware (mainly as Servers and End User Electronics) which provide the technical supporting platforms on which all the organisation operates. The other vital assets included the software Revenue Management System (RMS) used to process and manage business information processes, Organisation Reputation (intangible asset) which determines whether its business contracts be maintained or terminated in subsequent years and the Human Resource (staff) that are pillar in the operational success of the organization and are also the weakest link in the security chain.

### 3.4 Risk register

Using Qualitative Risk Analysis approach, this Section presents the detailed Risk Register framework for the organisation, highlighting assets threats, vulnerabilities and risks impact. A Threat is a potential cause of an incident that may result in harm to a system or organisation; Vulnerability is a weakness of an asset or group of assets that may be exploited by one or more threats and Impact is the result of Information Security incident, caused by a threat which affects assets [20] [1] [4]. A risk assessment approach is used to identify vulnerabilities, threats and impacts to which security controls will be applied.

Because of the logical simplicity and wide adaptability across several organisations, we have adopted the Asset Values and Impacts Matrix (TABLE. 1) to get a qualitative value of assets affected by risk incidents; Threat and Vulnerability Likelihood Table (TABLE .2) to quantify threat occurrences possibilities and quantify the degree of vulnerability. It should be noted that the weighing and values/figures in these tables are not fixed based on any known standard but rather on the logical significance of the interpretation as deemed fit by the organisation. In this report, the asset owners are responsible for accurately collecting the threats and vulnerabilities of every asset and the respective occurrences likelihood potentials whereas asset owners and suppliers are responsible for the value of the assets to be used in the Risks Register generated in *TABLE .3* below.

| LIKELIHOOD | DESCRIPTION | INTERPRETATION |
|---|---|---|
| 1 | Negligible | Once every 1000 years or less |
| 2 | Extremely Unlikely | Once every 200 years |
| 3 | Very Unlikely | Once every 50 years |
| 4 | Unlikely | Once every 20 years |
| 5 | Feasible | Once every 5 years |
| 6 | Probable | Annually |
| 7 | Very Probable | Quarterly |
| 8 | Expected | Monthly |
| 9 | Confidently Expected | Weekly |
| 10 | Certain | Daily |

TABLE.2 Threats and Vulnerability Likelihood Table

| RISK LINE | ID No. | Assets | Owner | Values (A) | Threats | Likelihood (T) | Vulnerabilities | Likelihood (V) | Risks (A*T*V) |
|---|---|---|---|---|---|---|---|---|---|
| ABOVE RISK APPETITE | 16 | Electronic Data | CIO | 5 | Human error | 8 | Mental Stress | 9 | 360 |
| | 18 | Electronic Data | CIO | 5 | Human error | 8 | Employees Physical fatigue | 9 | 360 |
| | 17 | Electronic Data | CIO | 5 | Human error | 8 | Fear to Consult | 8 | 320 |
| | 14 | Electronic Data | CIO | 5 | SQL Injections | 6 | Execution of malfunctioned queries irrespective of warnings | 9 | 270 |
| | 15 | Electronic Data | CIO | 5 | SQL Injections | 6 | Outdated DBMS | 9 | 270 |
| | 2 | Reputation | CEO | 5 | Data breach | 6 | Existence of backdoor | 8 | 240 |
| | 7 | Ruputation | CEO | 5 | Fraud | 6 | Disgruntled employees | 8 | 240 |
| | 13 | Electronic Data | CIO | 5 | SQL Injections | 6 | Bad coding Designs | 8 | 240 |
| | 1 | Reputation | CEO | 5 | Data breach | 6 | Outdated Security Software | 7 | 210 |
| | 3 | Reputation | CEO | 5 | Data breach | 6 | Reliance of reputation on trust | 7 | 210 |
| | 25 | Revenue Management System | CIO | 3 | Cross-Site Scripting | 7 | Susceptibility to Malicious code | 10 | 210 |
| | 27 | Revenue Management System | CIO | 3 | Cross-Site Scripting | 7 | User input support through input field | 10 | 210 |
| | 4 | Reputation | CEO | 5 | Misuse of resources | 6 | Staff dishonesty | 6 | 180 |
| | 5 | Reputation | CEO | 5 | Misuse of resources | 6 | Unclear resources utilization guidelines | 6 | 180 |
| | 6 | Reputation | CEO | 5 | Misuse of resources | 6 | Poor resource management | 6 | 180 |
| | 8 | Ruputation | CEO | 5 | Fraud | 6 | Unclear System Access Clearance guidelines | 6 | 180 |
| | 9 | Ruputation | CEO | 5 | Fraud | 6 | Staff dishonesty (Integrity loss or failure) | 6 | 180 |
| | 11 | Electronic Data | CIO | 5 | Data theft | 6 | Disgruntled employees | 6 | 180 |
| | 12 | Electronic Data | CIO | 5 | Data theft | 6 | Disposal of Storage media without proper erasure | 6 | 180 |
| | 26 | Revenue Management System | CIO | 3 | Cross-Site Scripting | 7 | Support for various unsafe scripting technologies. | 8 | 168 |
| | 34 | IT Hardware | CIO | 2 | Power interruptions | 9 | Inability to operate without power supply | 9 | 162 |
| | 36 | IT Hardware | CIO | 2 | Power interruptions | 9 | Fluctuating power supply (Unstable Power Supply) | 9 | 162 |
| | 37 | Staff | COO | 2 | Social engineering | 9 | Human tendency to be gullible (getting something for nothing) | 9 | 162 |
| | 38 | Staff | COO | 2 | Social engineering | 9 | Inclination for immediate gratification | 9 | 162 |
| RISK APPETITE | | | | | | | | | 150 |
| BELOW RISK APPETITE | 39 | Staff | COO | 2 | Social engineering | 9 | Inclination to Improved status gain | 9 | 162 |
| | 23 | Revenue Management System | CIO | 3 | Stack-Overflow attacks | 6 | Weaknesses in the programming language used to develop the systems | 8 | 144 |
| | 24 | Revenue Management System | CIO | 3 | Stack-Overflow attacks | 6 | Zero-day vulnerabilities | 8 | 144 |
| | 35 | IT Hardware | CIO | 2 | Power interruptions | 9 | Inability of backup power (UPS) to sustain long hours | 8 | 144 |
| | 19 | Revenue Management System | CIO | 3 | Denial Of Services | 6 | Low Memory Resources | 6 | 108 |
| | 20 | Revenue Management System | CIO | 3 | Denial Of Services | 6 | Limited Bandwidth | 6 | 108 |
| | 21 | Revenue Management System | CIO | 3 | Denial Of Services | 6 | Protocols in use such as Telnet | 6 | 108 |
| | 22 | Revenue Management System | CIO | 3 | Stack-Overflow attacks | 6 | Bad coding habits | 6 | 108 |
| | 42 | Staff | COO | 2 | Illness (Health) | 7 | Illnesses due to change of weather | 7 | 98 |
| | 10 | Electronic Data | CIO | 5 | Data theft | 6 | Breaching legal requirements | 3 | 90 |
| | 28 | IT Hardware | CIO | 2 | Heat | 6 | Susceptibility of Equipments to temperature variations | 7 | 84 |
| | 30 | IT Hardware | CIO | 2 | Heat | 6 | Susceptibility of Solder joints to melt at high temperatures | 7 | 84 |
| | 40 | Staff | COO | 2 | Illness (Health) | 7 | Weak immune systems due to genetic variations | 6 | 84 |
| | 41 | Staff | COO | 2 | Illness (Health) | 7 | Incomplete Immunisation to Common Diseases | 6 | 84 |
| | 43 | Staff | COO | 2 | Accidents | 6 | Ignorance to Precautions | 7 | 84 |
| | 44 | Staff | COO | 2 | Accidents | 6 | General carelessness and forgetfulness | 7 | 84 |
| | 45 | Staff | COO | 2 | Accidents | 6 | Tendency to take risks, being fearless | 7 | 84 |
| | 29 | IT Hardware | CIO | 2 | Heat | 6 | Susceptibility of Processor Chips to melt at high temperatures | 6 | 72 |
| | 31 | IT Hardware | CIO | 2 | Humidity | 4 | Non-water resistant Equipments | 8 | 64 |
| | 33 | IT Hardware | CIO | 2 | Humidity | 4 | Susceptibility to short circuiting due to water/moisture contact | 7 | 56 |
| | 32 | IT Hardware | CIO | 2 | Humidity | 4 | Susceptibility to corrosion due to rusting | 5 | 40 |

TABLE.3 Risk Assessment Register (the different assets in colour code)

## 3.5 Analysis

### 3.5.1 Risk Appetite

Risk appetite is the amount and type of risk that an organization is prepared to pursue, retain or take [21]. The organisation management establishes the risk appetite.

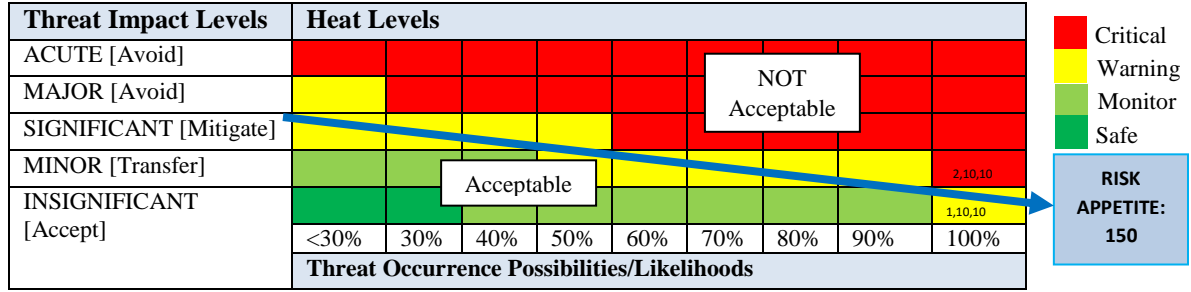

TABLE.4 Risk Register heat map

From the risk register in the *TABLE.3*, the Heat map (TABLE.4) is derived which is then shared with management to determine the Risk Appetite. For this organisation, the risk appetite line is set midway between (asset, threat, vulnerability) values (1, 10, 10) and (2, 10, 10) which respond to risks equal to 100 and 200 respectively. Our Risk appetite is therefore 150. Any threat above this appetite is not acceptable and must be avoided or controlled whereas any risks below 150 appetite line can be tolerated but subjected to handling guidelines of the organisation as specified in the ISM policies. Threats categorised as acute, major and critical (RED) have to be avoided and eliminated. Significantly harmful threats (YELLOW) are to be reduced by mitigation. Minor threats (GREEN) should be transferred by collaborating with an insurer whilst accepted insignificant threats need to be monitored.

### 3.5.2 Impact Controls

Controls are those activities that are taken to manage identified risks [22]. A control or countermeasure is placed to modify the risks and to counter threats. Controls can be used to prevent, deter, deflect, mitigate, detect and recover from a security breach, incident or compromise. The possibility for harm to is called risk [23]. ISO/IEC 27001 determines controls by design or from other sources. Some of the sources are the code of practice and industry standards. Although not all institutions can afford to purchase, install, operate and maintain expensive security controls and related systems, the decisions on security controls to be applied have to balance considerations of security risk and resource constraints. The identified control sets will change risks identified in the risk assessment process. They will be able to work together to reduce risks. For the implemented controls, it is vital to check whether new controls conflict with existing controls and whether there is any redundancy or how effectively they will work together.

The risk register in *TABLE.3* identifies Electronic Data, Reputation, Revenue Management System and Staff as high-risk assets. The control mechanisms to be applied in this assessment will aim at achieving the highest preventive techniques/controls of the respective threats. We shall adopt all three categories of control measures that include administrative, Technical and physical. Combination of specific information security policies in line with ISO270001 and NIST, elaborate monitoring, preventive, corrective, deterrent and discovery technical tools/measures and physical security techniques are best suited for effective controls. Therefore for effective information security adopting Least Privilege, Defense in Depth and separation of duties security principles approach, we incorporate several controls measures proposed below to eliminate, reduce, transfer or let the risks be accepted.

**Least Privilege Security:** no communications or activities to be permitted unless there is a need [3].
**Defence in depth Security:** use of multiple security techniques or layers of control to reduce exposure if one security control is compromised [3].
**Separation of Duties:** minimise errors and make it more difficult to exploit access privileges for personal gain [3].

**Administrative Controls**

Use of "soft controls" like policies, change control, user registration, visitors logs, punishment for failure to comply, roles, responsibilities and job descriptions. The following are specific Information Security Policies to ensure proper information security culture in the organisation: Encryption policy, Secure Data Transmission Policy, data lifecycle policy, Clean desk policy, Data breach policy, Digital signature acceptance policy, Disaster recovery plan policy, Email policy (cooperate emails labelling according to origin for easy identification of phishing emails), Non-disclosure agreement for service providers, Ethics policy, Password protection and construction policy, Security response plan policy. Acceptable Use Policy (AUP) can be used to reduce risks regarding the organisation's resources particularly IT systems. These policies will ensure social engineering and human error control and the overall information security management in the organisation will be improved. The policies should be reviewed every two (2) years. Implement continues information security awareness and Policies training. As an administrative control should also set up guidelines for competent and skilled personnel hiring along with Pre-employment background checks to control hiring fraudulent incompetent staff. Eventually, adopt Data classification and labelling, Separation of duties and Job rotations.

**Technical Controls**

The organisation should implement secure architecture with updated Firewalls, intrusion detection (IDS)/intrusion prevention (IPS) and Data Leakage Prevention (DLP) systems, Virus scanners, patch management, antispyware, Demilitarized Zones to control against data breaches, SQL injections, cross-site scripting and data theft. Data encryption should be enforced to provide data confidentiality, role-based access and account management control to manage staff access to information, Virtual Private Networks to provide secure communication channels. Routine data backups and server images for fail recovery, audit logs for routine review abnormal behaviours in the systems.

**Physical Controls**

Use of mantraps and locks to equipment rooms, security guards and CCTV to provide deterrent and detective controls, Badges, biometric access and swipe cards for access control against unauthorised access to facilities and equipment. The installations and use of backup power generators and batteries will provide control on power disruptions that affect the IT hardware and physical security.

### 3.5.3 Compensating Controls

A compensating control is one that meets the intent of the standard by reducing the exposure to an acceptable level, but it may not match the prescribed control in the standard or policy [3]. It is common that control objectives may not be fully achieved; hence, risks of irregularities in the business may exist. Management must therefore evaluate the cost-benefit of implementing additional controls called compensating controls to reduce the risks further [24]. Compensating controls include the Administrative, Technical or Logical and Physical elements as already been highlighted above.

### 3.5.4 Residual Risks

This is the remaining risk exposure level after implementing the recommended controls. It is calculated the same way as elaborated in earlier sections above. We should now be able to provide the management with before and after snapshots of how the recommended controls will lower the risks to the organisation [3].

## 4. FUTURE EXPANSIONS

A further expansion of this project would worth considering the cost-to-benefit repercussion of implementing the stated controls. Thomas talks about a refined Facilitated Risk Analysis, and Assessment Process (FRAAP) formulated from the bedrock of ISO 17799 which explains an optimised risk assessment process that takes into consideration the cost effectiveness of suggested control measures [25]. He made it clear that the existence of a threat does not necessarily mean an organisation is at risk.

The ultimate goal for mitigating risks is to protect the organisation's resources while ensuring the business maintains its objectives and mission. Hence, there could be situations whereby suggested controls could become infeasible for the organisation. Implementing some controls could ironically be the threat that leads to a business downfall. Controls should be practicable as such that the end justifies the means. Quantifying the overall cost of implementing each of the stated controls against the organisation's capability to implement them would create a room to determine what controls should be alternated for a compensating control as discussed in section 3.3.3.

## 5. CONCLUSIONS

In summary, we identified five key organisation assets electronic data, IT hardware, organisation reputation, the RMS software and staffs. The main threats faced by these assets are human errors from staffs, SQL injections, data breaches, fraud, cross-site scripting, data theft, social engineering and power interruptions that are responsible for the most prominent information security risks in the organisation. Appropriate industry-standard administrative, technical and physical controls have been employed to prevent, detect, mitigate and reduce the exploitation of vulnerabilities detected with the assets.

One fundamental limitation of this report is the scope of assets used. It is very rare to have an organisation with only five key assets. We strongly recommend that ISM practitioners work hand in hand with all stakeholders to ensure accurate information for all assets and their vulnerabilities are obtained. However, it should be noted that from this report, one should be able to follow systematically the processes and technicalities undertaken in preparing an information security risk assessment and management report. Irrespective of the choice of organisation for this report, the same principles used in this report apply globally to all industry sectors.

In this organisation report, a qualitative approach was used to assess the risks. However, other organisations would prefer to adopt more quantitative approaches which use hard metrics, such as pounds or dollars. It will be common to find other terms like exposure factors, single loss expectancy, the annual rate of occurrences, annualised loss expectancy and total cost of ownership that all end up being used to address the same issue of risks management. On combining with Risk Analysis, the Total Cost of Ownership and Return on Investment calculations should factor into proper budgeting.